\documentclass[12pt,a4paper]{article}
\usepackage[utf8]{inputenc}
\usepackage{amsmath,ulem} 
\usepackage[mathscr]{eucal}
\usepackage{color}
\usepackage[dvipsnames]{xcolor}
\input{epsf} 
\usepackage[numbers,sort&compress]{natbib}
\usepackage{setspace}
\newcommand\as{\alpha_{\mathrm{S}}} 
\newcommand\f[2]{\frac{#1}{#2}}

\def\beq{\begin{equation}} 
\def\eeq{\end{equation}} 
\def\beeq{\begin{eqnarray}} 
\def\eeeq{\end{eqnarray}} 
 
\def\to{\rightarrow} 
\def\nn{\nonumber}

\def\ms{${\overline {\rm MS}}$} 
\def\msbar{{\overline {\rm MS}}} 
 
\def\bnot{\beta_0}

\def\GE{\gamma_E}
\def\ep{\epsilon}

\def\At{{\widetilde {\mathscr A}}_{i}}
\def\A{{{\mathscr A}}_{i}}
\def\ATt{{\widetilde {\mathscr A}}_{T,i}}
\def\AT{{{\mathscr A}}_{T,i}}
\def\Aot{{\widetilde {\mathscr A}}_{0,i}}
\def\Ao{{{\mathscr A}}_{0,i}}

\def\cf{{\cal F}}
\def\Aft{{\widetilde {\mathscr A}}_{\cf,i}}

\def\Af{{{\mathscr A}}_{\cf,i}}

\def\Afft{{\widetilde {\mathscr A}}_{\cf/\cf(1),i}}
\def\Afto{{\widetilde {\mathscr A}}_{\cf_1,i}}
\def\Aftt{{\widetilde {\mathscr A}}_{\cf_2,i}}
\def\Af{ {\mathscr A }_{\cf,i}}
\def\Afone{ {\mathscr A }_{\cf_1,i}}
\def\Aftwo{ {\mathscr A }_{\cf_2,i}}
\def\Afp{ {\mathscr A }_{\cf=t^p,i}}

\def\AngInt{\Omega}

\def\M{{\cal M}}
\def\cp{{\cal P}}

\usepackage[hidelinks]{hyperref}
\usepackage{footnotebackref}

\begin{document} 
\begin{titlepage}
\begin{flushright}
TIF-UNIMI-2023-26,
ZU-TH 52/23,
PSI-PR-23-36
\end{flushright}

\vspace*{0.4cm}

\begin{center}
{\Large \bf Soft-gluon effective coupling: perturbative results \\[1ex] and the large-$\mathbf{n_F}$ limit to all orders}
\end{center}

\par \vspace{2mm}
\begin{center}
{\bf Stefano Catani${}^{(a)}$, 
Daniel de Florian${}^{(b)}$, Simone Devoto${}^{(c)}$,\\[0.3cm]
Massimiliano Grazzini${}^{(d)}$
and 
Javier Mazzitelli${}^{(e)}$}

\vspace{5mm}

${}^{(a)}$INFN, Sezione di Firenze and
Dipartimento di Fisica e Astronomia,\\ 
Universit\`a di Firenze,
I-50019 Sesto Fiorentino, Florence, Italy

${}^{(b)}$
International Center for Advanced Studies (ICAS) and ICIFI, ECyT-UNSAM,\\
25 de Mayo y Francia, (1650) Buenos Aires, Argentina

${}^{(c)}$Dipartimento di Fisica ``Aldo Pontremoli'', University of Milano and INFN, Sezione di Milano, I-20133 Milano, Italy

${}^{(d)}$Physik Institut, Universit\"at Z\"urich, CH-8057 Z\"urich, Switzerland

${}^{(e)}$Paul Scherrer Institut, CH-5232 Villigen PSI, Switzerland

\vspace{5mm}

\end{center}

\par \vspace{2mm}
\begin{center} {\large \bf Abstract} 

\end{center}
\begin{quote}
\pretolerance 10000

We consider extensions of the soft-gluon effective coupling that generalize the Catani--Marchesini--Webber (CMW) coupling 
in the context of soft-gluon resummation
beyond the next-to-leading logarithmic accuracy. Starting from the probability density of correlated soft emission in $d$ dimensions
we introduce a class of soft couplings relevant for resummed QCD calculations of
hard-scattering observables. We show that 
at the conformal point, 
where the $d$-dimensional QCD $\beta$ function vanishes, all these effective couplings are equal
and they are also equal to the cusp anomalous dimension.
We present explicit results in $d$ dimensions for	the soft-emission probability density and the	soft couplings at the second-order in the QCD coupling $\as$.
In $d=4$ dimensions we obtain the explicit relation between the soft couplings at ${\cal O}(\as^3).$
Finally, we study the structure of the soft coupling in the large-$n_F$ limit and we present
explicit expressions to all orders in perturbation theory. We also check that, at the conformal point, our large-$n_F$ results agree  with
the known result of the cusp	anomalous dimension.

\end{quote}

\vspace*{\fill}
\begin{flushleft}
September 2023
\end{flushleft}
\end{titlepage}

\setcounter{footnote}{1}
\renewcommand{\thefootnote}{\fnsymbol{footnote}}

\hrule
\renewcommand{\contentsname}{\normalsize Contents}
\tableofcontents{}

\vspace*{0.6cm}
\hrule
\vspace*{0.6cm}

\section{Introduction}

The QCD computations of hard-scattering processes can lead to large
contributions at each order of the perturbative expansion in the QCD
coupling $\as$.
Large logarithmic contributions always appear in the case of hard-scattering
observables
that are evaluated in kinematical regions close to the exclusive boundary of
the phase space. These large logarithms are produced by the emission of soft
and collinear partons, since hard emission is kinematically suppressed near
to the phase space boundary.

For a wide class of observables, logarithmic terms due to soft and collinear
emission
can be resummed to all perturbative orders in exponentiated form
(see, e.g., the reviews in Refs.~\cite{Luisoni:2015xha, Becher:2014oda} and
references therein). The exponentiated form can then be sistematically organized
and computed in terms of
leading logarithmic (LL) terms, next-to-leading logarithmic (NLL) terms,
next-to-next-to-leading logarithmic (NNLL) terms and so forth.
A relevant feature of these resummed expressions is that they have a quite
general structure with a high degree of universality and a `minimal'
dependence on the
hard-scattering process and on the specific observable to be treated.
In particular, the `dominant' (soft {\it and} collinear) part of the
logarithmic contributions is resummed and embodied in a `generalized'
Sudakov form factor,
whose kernel is perturbatively computable order-by-order in $\as$.

Known resummed results up to NLL accuracy show that the NLL kernel of the Sudakov form factor is entirely and
universally controlled through the use of the QCD coupling $\as^{\rm CMW}$ \cite{Catani:1990rr}
in the Catani--Marchesini--Webber (CMW) scheme
(or bremsstrahlung scheme). More precisely, the intensity $\A^{\rm CMW}$ of soft-gluon radiation from the hard-scattering parton $i$ $(i=q,{\bar q},g)$
in the Sudakov form factor at NLL accuracy is
\beq
\label{cmw}
\A^{\rm CMW}(\as(k_T^2))= C_i \,\f{\as^{\rm CMW}(k_T^2)}{\pi}
= C_i \,\f{\as(k_T^2)}{\pi}\left(1+\f{\as(k_T^2)}{2\pi} K\right)\;,
\eeq
where $C_i$ is the Casimir coefficient of the parton $i$
($C_i=C_A$ if $i=g$ and $C_i=C_F$ if $i=q, {\bar q})$,
$\as$ is the QCD running coupling in the \ms\ renormalization scheme,
$\as^{\rm CMW}$ is the CMW coupling \cite{Catani:1990rr} and
the coefficient $K$ depends on the number $n_F$ of quark flavours:
\beq
\label{kcoef}
K=\left(\f{67}{18}-\f{\pi^2}{6}\right) C_A -\f{5}{9}  n_F \;\;.
\eeq
The CMW coupling $\as^{\rm CMW}(k_T^2)$ has the meaning of an effective (physical) coupling for {\it inclusive}
radiation of soft {\it and} collinear partons with total transverse momentum $k_T$. 

The NLL universality of the CMW coupling is evident 
in the context of threshold resummation 
\cite{Sterman:1986aj, Catani:1989ne, Catani:1990rp},
transverse-momentum resummation
\cite{Collins:1984kg, Kodaira:1981nh, Catani:2000vq,Becher:2009qa}
and event shape resummation 
\cite{Catani:1992ua, Dasgupta:2002dc}.
Such universality is exploited for process-independent and observable-independent formulations of NLL resummations
\cite{Banfi:2003je, Bonciani:2003nt, Banfi:2004yd}.
The CMW coupling is also used to achieve NLL accuracy
in parton shower algorithms
\cite{Buckley:2011ms,Campbell:2022qmc,Bewick:2019rbu,
Dasgupta:2020fwr, Herren:2022jej}
for Monte Carlo event generators. 

Definitions of soft-gluon effective couplings to all perturbative orders
were introduced in Refs.~\cite{Banfi:2018mcq} and \cite{Catani:2019rvy}.  
These effective couplings
generalize the CMW coupling beyond the NLL level.

Reference \cite{Banfi:2018mcq} defined the effective coupling 
$\AT(\as)$ (according to the notation of Ref.~\cite{Catani:2019rvy}),
presented its explicit expression at ${\cal O}(\as^3)$ and discussed its use
in the context of NNLL resummations for two-jet observables in 
$e^+e^-$ annihilation 
\cite{Banfi:2014sua}.

Reference  \cite{Catani:2019rvy} observed that there is no unique extension
of the CMW coupling beyond NLL accuracy.  Two effective couplings were
considered in Ref.~\cite{Catani:2019rvy}: the couplings $\AT(\as)$
(as in Ref.~\cite{Banfi:2018mcq})
and $\Ao(\as)$. 
Moreover,  Ref.~\cite{Catani:2019rvy} considered the generalized $d$-dimensional
($d=4-2\epsilon$ is the number of space-time dimensions) extensions,
$\ATt(\as;\ep)$ and $\Aot(\as;\ep)$, of these effective couplings. 
A remarkable  result of  Ref.~\cite{Catani:2019rvy} is that the $d$-dimensional
couplings $\ATt(\as;\ep)$ and $\Aot(\as;\ep)$ are equal at the conformal point
$\ep=\beta(\as)$ ($\beta(\as)$  is the four-dimensional QCD $\beta$ function)
and equal to the cusp anomalous dimension \cite{Korchemsky:1988si,Moch:2004pa,Vogt:2004mw,
Korchemsky:1987wg,
Catani:1998bh, Aybat:2006mz, Gardi:2009qi, Becher:2009qa, Almelid:2015jia}.
Considering the four-dimensional couplings, 
Ref.~\cite{Catani:2019rvy} explicitly computed 
$\AT(\as)$ at ${\cal O}(\as^3)$  (confirming the result in Ref.~\cite{Banfi:2018mcq})
and $\Ao(\as)$ at ${\cal O}(\as^3)$ and  ${\cal O}(\as^4)$
(in terms of the cusp anomalous dimension at  ${\cal O}(\as^4)$ \cite{Henn:2019swt, vonManteuffel:2020vjv}).

The purpose of this paper is to further study the perturbative features of the soft-gluon effective coupling.
In $d$ dimensions we introduce an entire class of 
effective couplings, which includes the couplings
$\ATt(\as;\ep)$ and $\Aot(\as;\ep)$
of Ref.~\cite{Catani:2019rvy}.
Each effective coupling within this class is specified
by a scale that depends on the transverse momentum
and transverse mass of the inclusive soft radiation.
We show that all these effective couplings are equal
at the conformal point and, consequently \cite{Catani:2019rvy}, they are equal to the cusp anomalous dimension. 

We compute the effective couplings at ${\cal O}(\as^2)$
for arbitrary values of $d$
(of course, if $d=4$, the result at ${\cal O}(\as^2)$
is equal to the CMW coupling).
In $d=4$ we present the explicit relation between the effective couplings at ${\cal O}(\as^3)$. Considering the limit of a large value of the number $n_F$ of quark flavours, we explicitly evaluate the soft-gluon effective couplings to all orders in $\as$.

The paper is organised as follows. In Sect.~\ref{sec:def} we introduce the probability density of correlated soft emission and the soft-gluon effective coupling. In Sect.~\ref{sec:pert} we present the explicit computation of the correlated emission and of the soft coupling at ${\cal O}(\as^2)$. Some general features of
the soft-gluon effective coupling are discussed in Sect.~\ref{s:general}, while in Sect.~\ref{sec:largenf} we present our all-order results in the large-$n_F$ limit. Our results are summarised in Sect.~\ref{sec:summa}.

\section{Soft factorisation, correlated 
emission
and soft-gluon effective couplings}
\label{sec:def}

The perturbative calculations that we carry out in the paper are performed by
using analytic continuation in $d=4 -2\ep$ space time dimensions to regularise
both ultraviolet (UV) and infrared (IR) divergences.
Specifically, we use the customary scheme of 
conventional dimensional regularisation (CDR) \cite{Bollini:1972ui,Ashmore:1972uj,Cicuta:1972jf,Gastmans:1973uv} with $d-2$ spin
polarisation states for on-shell gluons and 2 spin polarisation states for on-shell massless quarks and antiquarks. The UV divergences are removed by renormalisation
in the \ms\ scheme. IR divergences, which are encountered at intermediate stages of the calculations, cancel in the computation of IR and collinear-safe quantities.

The QCD bare coupling $\as^u$ and the renormalised running coupling $\as(\mu_R^2)$
in the \ms\ scheme are related by the following standard definition:
\begin{equation}
\label{asren}
\as^u \,\mu_0^{2\ep}S_\ep = \as(\mu_R^2) \,\mu_R^{2\ep} \;Z(\as(\mu_R^2);\ep)
\;\;,
\end{equation}
\beq
\label{sep}
S_\ep \equiv (4\pi)^\ep \,e^{-\ep\gamma_E}
\eeq
where $\mu_0$ is the dimensional regularisation scale, $\mu_R$ is the renormalisation scale and $\gamma_E$ is the Euler number. 
The renormalisation function of the coupling is
\begin{equation}
Z(\as,\ep)=1-\as\f{\beta_0}{\ep}+\as^2
\left(\f{\beta_0^2}{\ep^2}-\f{\beta_1}{2\ep}\right)+{\cal O}(\as^3)
\;\;,
\end{equation}
where $\beta_0$ and $\beta_1$ are the first two perturbative coefficients
of the QCD $\beta$-function $\beta(\as)$:
\beq
\label{betaf}
\beta(\as) = - \beta_0 \,\as -  \beta_1 \,\as^2 +  {\cal O}(\as^3) \;\;,
\eeq
\begin{align}
\label{betacoeff}
&12\pi \,\beta_0=11C_A-4 T_R n_F \;\;,~~~~~~~~~~~~~24\pi^2\beta_1=17C_A^2-10 C_A T_R n_F-6 C_F T_R n_F\; ,
\end{align}
and $n_F$ is the number of quark flavours. The colour coefficients in $SU(N_c)$ QCD with $N_c$ colours are $C_A=N_c$, $C_F=(N_c^2-1)/(2N_c)$ and $T_R=1/2$.
The scale dependence of the renormalised coupling in $d$ dimensions
is controlled by the following
evolution equation 
\beq
\label{scaleev}
\frac{d \ln \as(\mu_R^2)}{d \ln \mu_R^2} = - \ep + \beta(\as(\mu_R^2)) \;\;,
\eeq
and $\beta(\as) - \ep$ is the 
$d$-dimensional $\beta$-function.

The all-order soft-gluon effective coupling is formally defined \cite{Banfi:2018mcq,
Catani:2019rvy} by considering multiple soft-parton radiation in the simplest class
of hard-scattering processes, namely, the processes that involve only two massless
hard partons. Owing to colour conservation in the soft limit, the two hard partons have
colour charge in complex-conjugate representations and they 
can be either a $q{\bar q}$ pair ($i=q$) or two gluons ($i=g$).
This class of processess is used  to define the function
$w_i(k;\ep)$ that gives the `probability'\footnote{Note that this `probability' is not positive definite since it refers to
the correlation part of the 
(positive definite) 
total emission probability.} 
of {\it correlated} emission (including the corresponding virtual corrections) of an arbitrary number of soft partons with total
momentum $k$. The correlated probability $w_i(k;\ep)$ is a quantity that is IR and collinear safe. 
The soft-gluon effective coupling is obtained by properly integrating $w_i(k;\ep)$
over the soft momentum $k$.
In the following we give details on this procedure starting from soft-factorisation formulae of QCD scattering amplitudes.

We consider a generic scattering process whose external particles are two QCD hard partons $i$ and $\bar i$, soft partons and additional colourless particles.
The corresponding scattering amplitude with $N$ external soft partons is denoted as 
$\M_{i\bar i}(p_1,p_2; k_1, \dots, k_N)$, where $p_1$ and $p_2$ are the 
momenta of the hard partons and $k_1, \dots, k_N$ are the momenta of the soft partons.
Hard and soft partons are massless.
In the kinematical region where all the momenta $k_\ell$ ($\ell= 1, \dots, N$)
simultaneously vanish the squared amplitude $| \M_{i\bar i} |^2$ (summed over the colours and spins of the external particles) is singular. The dominant singular behaviour is given by the following soft-factorisation formula:
\beq
\label{ff}
| \M_{i\bar i}(p_1,p_2; k_1, \dots k_N) |^2 \simeq 
| \M_{i\bar i}(p_1,p_2) |^2 \;| J_i(k_1,\cdots,k_N) |^{2} \;,
\eeq
where $J_i$ is the soft-parton current and $| J_i |^{2}$ is the corresponding squared current. The factorisation formula in Eq.~(\ref{ff}) is valid
\cite{Bern:1999ry, Catani:2000pi, Feige:2014wja} at arbitrary orders in the loop expansion of the scattering amplitude and, correspondingly, $J_i$ is the 
{\it all-order} soft current. In particular, the squared current $| J_i |^{2}$
(analogously to $| \M_{i\bar i} |^2$)
can be expressed in terms of the renormalised coupling and it can be expanded in powers of $\as(\mu_R^2)$. The coefficients of this perturbative expansion
have no UV divergences, but they still have IR divergences (in the form of $\ep$ poles). These IR divergences cancel by combining the squared current contributions in the computation of IR and collinear-safe quantities.

We recall that the soft-factorisation formula for generic scattering amplitudes 
leads to colour correlations between the hard partons. In the case of two sole hard partons, colour correlations are effectively removed since there is only {\it one}
colour singlet configuration of the two hard partons. Consequently, soft factorisation takes the form in Eq.~(\ref{ff}) (see, e.g., Sect.~7.1 in 
Ref.~\cite{Catani:2019nqv}),
where $| J_i |^{2}$ is a $c$-number function (rather than a colour matrix operator).

The squared current $| J_i(k_1,\cdots,k_N) |^{2}$ depends on the soft momenta and on the momenta and colour representation of the two hard partons. The hard-parton dependence is simply denoted by the subscript $i$ in $J_i$. 

The soft limit $k_\ell \to 0$ is insensitive to the actual size of the energies of the hard partons. This implies that the squared current $| J_i |^{2}$ is invariant under the rescaling transformation $p_1 \to \xi_1 p_1$, $p_2 \to \xi_2 p_2$ ($\xi_1$ and $\xi_2$ are arbitrary parameters) of the hard-parton momenta.\footnote{Such invariance is evident
by using the eikonal approximation for the computation of the soft current (see, e.g., Ref.~\cite{Catani:1999ss}).} Consequently, $| J_i |^{2}$ is invariant under longitudinal boosts along the direction of the momenta of the two hard partons in their centre--of--mass frame.

Owing to these invariance properties of the soft current, it is convenient to introduce kinematical variables that are longitudinally boost invariant. For an arbitrary (i.e., off-shell) $d$-dimensional momentum $k^\nu$ we define
\beq
\label{kin}
m_T^2 = \frac{2 (p_1 \cdot k) (p_2 \cdot k)}{p_1 \cdot p_2} \;, 
\quad
k_T^2 = m_T^2 - k^2 \;,  \quad
y = \frac{1}{2} \ln\frac{p_2 \cdot k}{p_1 \cdot k} \;\;,
\eeq
where $m_T$ is the transverse mass, $y$ is the rapidity and $k_T$ is the 
(modulus of the) transverse momentum with respect to the direction of the hard partons.

At the leading order (LO) in QCD perturbation theory, we have to consider only the radiation of a single soft gluon (with momentum $k$) and the corresponding LO
squared current $| J_i(k) |^{2}_{(\text{LO})}$.
The LO differential probability
$d\cp_{i, \text{LO}}$ of soft radiation is obtained by supplementing 
$| J_i(k) |^{2}_{(\text{LO})}$ with the corresponding $d$-dimensional phase space,
and we have
\beq
\label{plo}
d\cp_{i, \text{LO}} = \frac{d^dk}{(2\pi)^{d-1}} \;\delta_+(k^2) 
\;| J_i(k) |^{2}_{(\text{LO})} \equiv dy \, dm_T^2 \, dk_T^2 \; \frac{1}{2} \;
w_{i, \text{LO}}(k;\ep) \;\;.
\eeq 
In the right-hand side of Eq.~(\ref{plo}) we have introduced the kinematical variables of Eq.~(\ref{kin}) and we have defined the LO probability density $w_{i, \text{LO}}$,
which embodies the complete $d$-dimensional dependence on the phase space.
Since the phase space factor is
\beq
\label{nep}
d^dk = \frac{1}{2} \;dy \, dm_T^2 \, dk_T^2 \; N_{\ep}(k_T) \;\;,
\quad N_{\ep}(k_T) = \frac{\pi^{1-\ep} (k_T^2)^{-\ep}}{\Gamma(1-\ep)} \;\;,
\eeq
where $\Gamma(z)$ is the Euler gamma function, we have
\beq
\label{wlodef}
w_{i, \text{LO}}(k;\ep) = \frac{N_{\ep}(k_T)}{(2\pi)^{d-1}} 
\;| J_i(k) |^{2}_{(\text{LO})}  \;\delta_+(k^2) \;\;.
\eeq

The LO expression of the renormalised squared current is 
(see, e.g., Ref.~\cite{Catani:1999ss})
\beq
\label{jlo}
| J_i(k) |^{2}_{(\text{LO})} = 8\pi \; \frac{p_1 \cdot p_2}{p_1 \cdot k \;p_2 \cdot k}
\;(S_\ep)^{-1} \;C_i \;\as(\mu_R^2) \; (\mu_R^2)^{\ep} \;\;,
\eeq
where $S_\ep$ is defined in Eq.~(\ref{sep}) and $C_i$ is the Casimir coefficient of the radiating hard parton $i$ ($C_i=C_A$ if $i=g$, and $C_i=C_F$ if
$i=q \;{\text or}\;{\bar q}$).
Therefore, inserting Eq.~(\ref{jlo}) in  Eq.~(\ref{wlodef}), the expression of the LO probability density is
\beq
\label{wlo}
w_{i, \text{LO}}(k;\ep) = \frac{2}{m_T^2} \;\delta(m_T^2 - k_T^2) 
\;C_i \;\frac{\as(\mu_R^2)}{\pi} \;\left(\frac{\mu_R^2}{k_T^2} \right)^\ep \,c(\ep) \;,
\eeq
where
\beq
\label{c1}
c(\ep) \equiv \f{e^{\ep\GE}}{\Gamma(1-\ep)} = 1 - \f{\pi^2}{12}  \, \ep^2
- \f{1}{3} \,\zeta_3 \, \ep^3 + {\cal O}(\ep^4) \;\;,
\eeq
and $\zeta_k$ is the Riemann $\zeta$-function.
We note that $w_{i, \text{LO}}(k;\ep)$ does not depend on the rapidity of the soft momentum $k$. The rapidity independence is a consequence of the boost invariance properties of soft radiation 
and it persists at any perturbative orders. The density $w_{i}(k;\ep)$ depends on 
$m_T^2$ and $k_T^2$. The factor $1/m_T^2$ in the right-hand side of Eq.~(\ref{wlo})
simply follows from the eikonal factor 
$(p_1 \cdot k \;p_2 \cdot k/p_1 \cdot p_2)^{-1}$ in the squared current of 
Eq.~(\ref{jlo}). The $m_T$ and $k_T$ dependence of $w_{i, \text{LO}}(k;\ep)$
is directly constrained by the on-shell condition $\delta(k^2) = \delta(m_T^2 - k_T^2)$ for single soft-gluon emission. At higher perturbative orders the total momentum $k$ of the soft partons is not on-shell and, consequently, $w_{i}(k;\ep)$ has a non-trivial dependence on $m_T^2$ and $k_T^2$ (see Sect.~\ref{sec:pert}).

At the next-to-leading order (NLO) in QCD perturbation theory the probability of soft radiation receives contributions from single-parton and double-parton emissions.
The renormalised squared current $| J_i(k) |^{2}$ for single-gluon emission
has the following perturbative expansion:
\beq
\label{1gnlo}
| J_i(k) |^{2} = | J_i(k) |^{2}_{(\text{LO})} + | J_i(k) |^{2}_{(\text{NLO})} + \dots \;\;,
\eeq
where $| J_i(k) |^{2}_{(\text{LO})}$ is given in Eq.~(\ref{jlo}), the NLO term 
$| J_i(k) |^{2}_{(\text{NLO})}$ is obtained from the one-loop soft-gluon current
\cite{Bern:1999ry, Catani:2000pi} and the dots stand for higher-order terms
(the two-loop soft-gluon current is computed in Refs.~\cite{Li:2013lsa,Duhr:2013msa}).
In the case of double-parton radiation, we have to consider the squared current
$|J_i^{gg}(k_1,k_2)|^{2}_{(\text{LO})}$ for double-gluon emission and the squared current
$|J_i^{q\bar q}(k_1,k_2)|^{2}_{(\text{LO})}$ for the emission of a quark-antiquark pair of a given flavour. The renormalised LO expressions of these two-parton squared currents are directly proportional to the tree-level squared currents of Ref.~\cite{Catani:1999ss}, and we write
\beeq
\label{2glo}
|J_i^{gg}(k_1,k_2)|^{2}_{(\text{LO})} &=& | J_i(k_1) |^{2}_{(\text{LO})} 
\;| J_i(k_2) |^{2}_{(\text{LO})} +  W_i^{gg}(k_1,k_2)_{(\text{LO})} \;\;, \\
\label{qqlo}
|J_i^{q\bar q}(k_1,k_2)|^{2}_{(\text{LO})} &\equiv&  W_i^{q\bar q}(k_1,k_2)_{(\text{LO})}
\;\;.
\eeeq
In the right-hand side of Eq.~(\ref{2glo}), the first term is the contribution of the independent emission of the two soft gluons and, therefore, $W_i^{gg}(k_1,k_2)$
is the {\it irreducible-correlation} contribution to $|J_i^{q\bar q}(k_1,k_2)|^{2}$.
In Eq.~(\ref{qqlo}) the irreducible-correlation component $W_i^{q\bar q}(k_1,k_2)$
is equal to the entire squared current $|J_i^{q\bar q}(k_1,k_2)|^{2}$ since the soft quark and antiquark cannot be radiated independently (i.e., the soft current for emission of a quark or an antiquark vanishes).

We consider the differential probability $d\cp_i$ of {\it correlated} emission of an arbitrary number and type of soft partons. At fixed value of the total momentum $k$
of the soft partons, $d\cp_i$ can be expressed through the all-order probability density
$w_{i}(k;\ep)$, which generalizes the LO expressions in Eqs.~(\ref{plo}) and (\ref{wlodef}).
We write
\beq
\label{dpall}
d\cp_i =  dy \, dm_T^2 \, dk_T^2 \; \frac{1}{2} \;
w_i(k;\ep) \;\;,
\eeq 
where $w_i$ has the following perturbative expansion:
\beq
\label{wpert}
w_i(k;\ep) =
w_{i, \text{LO}}(k;\ep) + w_{i, \text{NLO}}(k;\ep) + \dots \;\;,
\eeq
and the dots denote higher-order terms.

The NLO contribution $w_{i, \text{NLO}}$ to the probability density $w_i$ in 
Eq.~(\ref{wpert}) is given in terms of the correlation component of the single-emission
and double-emission soft factors in Eqs.~(\ref{1gnlo})--(\ref{qqlo}). Specifically,
we have
\beeq
  \label{eq:web}
  w_{i,\text{NLO}}(k;\epsilon)&=&N_\ep(k_T)\bigg\{
  \frac{1}{(2\pi)^{d-1}} \delta_+(k^2) \;| J_i(k) |^{2}_{(\text{NLO})}
\\
  &&\hspace{-3.0cm} +
  \int
  \frac{d^d k_1}{(2\pi)^{d-1}} \delta_+(k_1^2)
  \frac{d^d k_2}{(2\pi)^{d-1}} \delta_+(k_2^2)
    \left(
    \frac{1}{2}\, W_i^{gg}(k_1,k_2)_{(\text{LO})} + n_F \,W_i^{q\bar q}(k_1,k_2)_{(\text{LO})}
    \right)
  \delta^{(d)}\left(k- k_1 -k_2\right)
  \bigg\} \;,\nn
  \eeeq
and the explicit computation of $w_{i,\text{NLO}}$ is presented in Sect.~\ref{sec:pert}.

The general all-order definition of the probability density $w_i(k;\ep)$ was given
in Ref.~\cite{Banfi:2018mcq}, where $w_i(k;\ep)$ was called {\it web function}. We have
\beeq
  \label{weball}
  w_i(k;\epsilon)&=&N_\ep(k_T)\bigg\{
  \frac{1}{(2\pi)^{d-1}} \delta_+(k^2) \;| J_i(k) |^{2}
\\
  &&\hspace{-2.2cm} +
   \sum_{N=2}^{+\infty} \;\;\sum_{\{a_1,\dots,a_N\}} S_{a_1,\dots,a_N}
  \int
  \left[ \prod_{i=1}^N \frac{d^d k_i}{(2\pi)^{d-1}} \delta_+(k_i^2) \right]
    \; W_i^{a_1 \dots a_N}(k_1,\dots,k_N) \;\;
  \delta^{(d)}(k- \sum_{j=1}^N k_j )
  \bigg\} \;,\nn
\eeeq
where $W_i^{a_1 \dots a_N}(k_1,\dots,k_N)$ denotes the {\it irreducible-correlation}
component of the squared current 
$| J_i^{a_1 \dots a_N}(k_1,\dots,k_N) |^{2}$ for emission of a set of $N$ partons with flavours $a_1,\dots,a_N$ ($a_i=g,q_f, {\bar q}_f$ and $q_f$ denotes a quark of
flavour $f$). At fixed soft-parton multiplicity $N$, Eq.~(\ref{weball}) involves the sum
over all parton configurations $\{a_1,\dots,a_N\}$ and $S_{a_1,\dots,a_N}$ is the customary
Bose symmetry factor for identical partons in each parton configurations
(e.g., $S_{a_1,\dots,a_N}=1/N!$ if all partons are gluons). The $N$-parton irreducible correlation $W_i^{a_1 \dots a_N}(k_1,\dots,k_N)$ is directly extracted from 
$| J_i^{a_1 \dots a_N}(k_1,\dots,k_N) |^{2}$ by subtracting the contributions from independent emission and from correlations of a lower number $M$ ($M < N$) of partons
(see, e.g., Ref.~\cite{Catani:2019nqv}
for explicit expressions with $N\leq 4$  and Eqs.~(8.1)--(8.5) therein for general $N$).

Note that $| J_i(k) |^{2}$ and $W_i^{a_1 \dots a_N}(k_1,\dots,k_N)$ in Eq.~(\ref{weball})
are all-order renormalised expressions, and they have their customary perturbative expansion
in terms of contributions at LO, NLO and so forth (see, e.g., Eq.~(\ref{1gnlo})).
The virtual radiative corrections embodied in $| J_i(k) |^{2}$ and $W_i^{a_1 \dots a_N}(k_1,\dots,k_N)$ lead to IR divergences and corresponding $\ep$ poles in the limit $\ep \to 0$.
However the probability density $w_i(k;\epsilon)$ is an infrared and collinear-safe quantity. Therefore, in Eq.~(\ref{weball}) the virtual IR divergences are cancelled by
the IR divergences 
that are produced by the integration over the momenta $k_i$ of the real-emission partons
in $W_i^{a_1 \dots a_N}(k_1,\dots,k_N)$. The cancellation of the IR divergences takes place
order-by-order in the perturbative expansion with respect to $\as(\mu_R^2)$, and at each perturbative order $w_i(k;\epsilon)$ is finite at $\ep=0$. Throughout the paper we consider
the general $d$-dimensional function $w_i(k;\epsilon)$, although it is well defined in the physical four-dimensional limit $\ep \to 0$.

The probability density $w_i(k;\epsilon)$ can be used to define soft-gluon effective couplings \cite{Banfi:2018mcq, Catani:2019rvy},
which generalise the NLO CMW coupling $\A^{\rm CMW}$ in Eq.~(\ref{cmw}).
As previously observed, $w_i(k;\epsilon)$ depends on two kinematical scales, namely,
the transverse momentum $k_T$ and the transverse mass $m_T$. The soft effective couplings
are obtained by integrating $w_i(k;\epsilon)$ with respect to one of these two
scales. Following Ref.~\cite{Catani:2019rvy}, we consider the $d$-dimensional
effective couplings $\ATt(\as;\ep)$ and $\Aot(\as;\ep)$, which are defined as 
\begin{align}
  \label{eq:defAT}
  \ATt(\as(\mu^2);\ep)&=\f{1}{2}\;\mu^2\int_0^\infty dm_T^2 \,dk_T^2 \;\delta(\mu^2-k_T^2) \;w_i(k;\ep) \;\;,\\
  \label{eq:defAo}
  \Aot(\as(\mu^2);\ep)&=\f{1}{2}\;\mu^2\int_0^\infty dm_T^2 \,dk_T^2 \;\delta(\mu^2-m_T^2) \;w_i(k;\ep) \;\;,
\end{align}
where $\ATt(\as;\ep=0)$ corresponds to the four-dimensional
effective coupling introduced in Ref.~\cite{Banfi:2018mcq}.

The definitions in Eqs.~(\ref{eq:defAT}) and (\ref{eq:defAo}) differ only in the
kinematical variable that is kept fixed in the integration procedure over $k$: 
$\ATt(\as(\mu^2);\ep)$ is defined at fixed value $k_T=\mu$ of the 
transverse momentum, while $\Aot(\as(\mu^2);\ep)$ is defined at fixed value
$m_T=\mu$ of the transverse mass. In this paper, we also propose a generalised version
of the couplings $\ATt$ and $\Aot$, which is introduced by keeping fixed a scale $\mu$
that depends on both $k_T$ and $m_T$. The generalised effective coupling $\Aft$ is defined
as follows:
\begin{equation}
\label{eq:softcoupling_f}
\Aft(\as(\mu^2);\ep)=\frac 12\;\mu^2\int_0^\infty \,dm_T^2\,dk_T^2 \;\delta\!\left(\mu^2-\frac{k_T^2}{\cf(k_T^2/m_T^2)}\right) \;w_i(k;\ep) \;\;,
\end{equation}
where $\cf(k_T^2/m_T^2)$ is a generic smooth function of $k_T^2/m_T^2$ that is equal to unity
in the limit $k_T \to m_T$ of on-shell radiation ($k^2 \to 0$),
namely, $\cf(k_T^2/m_T^2) =1$ at $k_T = m_T$ (see more comments
in Sect.~\ref{s:general}). The coupling in Eq.~(\ref{eq:softcoupling_f})
reduces to $\ATt$ and $\Aot$ for $\cf(k_T^2/m_T^2)=1$ and $\cf(k_T^2/m_T^2)=k_T^2/m_T^2$, respectively. In general, the effective coupling $\Aft(\as(\mu^2);\ep)$ measures the intensity
of {\it inclusive} correlated emission of soft partons at the momentum scale
$\mu^2=k_T^2/\cf(k_T^2/m_T^2)$.  

As mentioned in Ref.~\cite{Catani:2019rvy}, the effective coupling $\Aot$ has direct applications in the context of threshold resummation
\cite{Sterman:1986aj, Catani:1989ne, Catani:1990rp}.
Similarly, the coupling $\ATt$ is directly relevant for transverse-momentum resummation
\cite{Collins:1984kg, Kodaira:1981nh, Catani:2000vq,Becher:2010tm}.
By properly choosing the function $\cf(k_T^2/m_T^2)$ in Eq.~(\ref{eq:softcoupling_f}),
one can introduce soft effective couplings $\Aft$ that can be useful for resummed calculations of different classes of hard-scattering observables. The generalised definition 
in Eq.~(\ref{eq:softcoupling_f}) is also relevant to highlight more formal properties of the soft-gluon effective couplings (see Sect.~\ref{s:general}).

In the right-hand side of Eqs.~(\ref{eq:defAT})--(\ref{eq:softcoupling_f}), the factor
$\mu^2$ is introduced for dimensional reasons, so that the effective 
coupling\footnote{We use the notation $\At$ without subscript $T, 0$ and $\cf$
if we consider features that refer to all the effective couplings $\ATt, \Aot$ and $\Aft$.} 
$\At$ is dimensionless.
In the definitions of
Eqs.~(\ref{eq:defAT})--(\ref{eq:softcoupling_f})
the renormalisation scale $\mu_R$ of the QCD coupling $\as$ is set to the value $\mu_R=\mu$.
Obviously, the soft coupling $\At(\as(\mu^2);\ep)$ 
is a renormalisation group invariant quantity, so that, at the perturbative level,
it can equivalently be expressed in terms of the running coupling $\as(\mu_R^2)$
and the ratio $\mu^2/\mu_R^2$
(i.e., $\as(\mu^2)$ can be re-expressed in terms of $\as(\mu_R^2)$ and $\mu^2/\mu_R^2$).

The effective coupling $\At$ in Eqs.~(\ref{eq:defAT})--(\ref{eq:softcoupling_f}) is defined
to all perturbative orders in an arbitrary number $d=4-2\ep$ of space-time dimensions.
The integration over $m_T$ and $k_T$ in Eqs.~(\ref{eq:defAT})--(\ref{eq:softcoupling_f})
is infrared and collinear safe and, consequently, the limit $\ep \to 0$ is finite and well defined. By simply setting $\ep=0$ in $\At$, we have
\begin{equation}
\A(\as)\equiv \At(\as;\ep=0) \;\;,
\end{equation}
which defines the soft-parton effective couplings $\AT(\as), \Ao(\as)$ and $\Af(\as)$
in the physical four-dimensional space-time.
The perturbative expansions of $\At$ and $\A$ 
are written as
\beq
\label{atpert}
\At(\as;\ep) = \sum_{n=1}^\infty \left(\frac{\as}{\pi}\right)^n \At^{(n)}(\ep)\;,
\quad \quad
\A(\as) = \sum_{n=1}^\infty \left(\frac{\as}{\pi}\right)^n \A^{(n)}\;,
\eeq
where the coefficient $\At^{(n)}(\ep)$ depends on $\ep$ and $\A^{(n)}=\At^{(n)}(\ep=0)$.

The LO coefficients $\At^{(1)}(\ep)$ and $\A^{(1)}$ are obtained by inserting the LO expression
$w_{i, \text{LO}}(k;\ep)$ of Eq.~(\ref{wlo}) in 
Eqs.~(\ref{eq:defAT})--(\ref{eq:softcoupling_f}), and we get
\beq
\label{a1coeff}
\Aft^{(1)}(\ep) =
C_i \; c(\ep) \;\;, \quad \quad \Af^{(1)} = 
C_i \equiv A_i^{(1)} \;\;,
\eeq
where $c(\ep)$ is given in Eq.~(\ref{c1}). We note that $\Aft^{(1)}(\ep)$ is completely independent of the specific form of the function $\cf(k_T^2/m_T^2)$ in 
Eq.~(\ref{eq:softcoupling_f}) (we recall that $\cf(k_T^2/m_T^2)=1$ at $k_T=m_T$).
This independence simply follows from the fact that the 
lowest-order contribution to $w_i(k;\ep)$ is proportional to $\delta(k^2)=
\delta(m_T^2 - k_T^2)$. Therefore, all the $d$-dimensional soft effective couplings that we are considering are exactly equal at the LO.
The $\ep$ dependence of $\At^{(1)}(\ep)$ in Eq.~(\ref{a1coeff}) is due to $c(\ep)$
and it is entirely of kinematical origin, since it arises from 
the $d$-dimensional phase space of the total soft momentum $k$.
In the four-dimensional case, the coefficient $\A^{(1)}$ is simply equal to the Casimir charge
$C_i$ of the radiating hard parton $i$.

The computation of the $d$-dimensional couplings $\At(\as;\ep)$ at NLO is presented in 
Sect.~\ref{sec:pert}.
We anticipate that, in the physical four-dimensional space time, the NLO soft coupling 
$\Af$ is also completely independent of the specific function $\cf$.
Therefore,
all the four-dimensional soft couplings $\A$ defined through Eqs.~(\ref{eq:defAT})--(\ref{eq:softcoupling_f})
are equal to the CMW coupling $\A^{\rm CMW}$ up to ${\cal O}(\as^2)$, and we have  
\begin{equation}
\label{af2}
\Af^{(2)} = A^{(2)}_i \;\;.
\end{equation}
where
\begin{equation}
\label{a2coef}
A^{(2)}_i = \frac{1}{2} \,C_i \;K =  \frac{1}{2} \,C_i \;\left[ \left(\f{67}{18}-\f{\pi^2}{6}\right) C_A -\f{10}{9}T_R n_F \right]\;.
\end{equation}
The origin of the NLO equality of the four-dimensional soft effective couplings  
is discussed in Sects.~\ref{sec:pert} and \ref{s:general}.

\section{Perturbative results}
\label{sec:pert}

The NLO probability density $w_{i,{\rm NLO}}$ receives contributions from soft single-gluon emission at the one-loop level and from the soft emission of two gluons and a quark-antiquark pair.   
The corresponding soft-current factors in Eq.~(\ref{eq:web}) can be found 
in Refs.~\cite{Catani:1999ss,Catani:2000pi}. We can cast them in the following form:
\begin{align}
\label{jnlo}
  | J_i(k) |^{2}_{(\text{NLO})} =& -\as(\mu_R^2)\left[
   \frac{c(2\ep)}{2c(\ep)}C_A \frac{\pi \cos(\pi\ep)}{\sin^2(\pi\ep)}  
  \left(
  \frac{\mu_R^2 (p_1 \cdot p_2)}{2 (p_1 \cdot k) (p_2 \cdot k)}
  \right)^\ep
  +\frac{\beta_0}{\ep}
  \right]| J_i(k) |^{2}_{(\text{LO})} \;,\\[1.4ex]
W_i^{q\bar q}(k_1,k_2)_{(\text{LO})}  =& \;  (4\pi\, \alpha_S(\mu_R^2) \, \mu_R^{2\ep} S_\ep^{-1})^2 \,C_i\, {\widehat J}_\mu(k_1+k_2) \Pi^{\mu\nu}(k_1,k_2)  {\widehat J}_\nu(k_1+k_2) \,, 
\label{wqq} \\[2ex]
W_i^{gg}(k_1,k_2)_{(\text{LO})}  =& \; (4\pi\, \alpha_S(\mu_R^2) \, \mu_R^{2\ep}S_\ep^{-1})^2 \, C_A\, C_i
\left[ -{\widehat J}_\mu(k_1+k_2) \widetilde\Pi^{\mu\nu}(k_1,k_2)  {\widehat J}_\nu(k_1+k_2) 
  +
   2\, \widetilde{\cal S}(k_1,k_2) \right] \,,
   \label{wgg}
\end{align}
where ${\widehat J}^\mu(k)$ is proportional to the LO current,
\beq
{\widehat J}^\mu(k)= \frac{p_1^\mu}{p_1 \cdot k} - \frac{p_2^\mu}{p_2 \cdot k} \;,
\label{jhat}\eeq
and we have defined 
\begin{align}
\label{piqq}
\Pi^{\mu\nu}(k_1,k_2)=&\;\frac{T_R}{(k_1\cdot k_2)^2}\left(-g^{\mu\nu} k_1\cdot k_2 +k_1^\mu k_2^\nu+k_1^\nu k_2^\mu\right)\,, \\
\widetilde \Pi^{\mu\nu}(k_1,k_2)=&-\frac {1}{(k_1 \cdot k_2)^2}\left(-4 g^{\mu\nu}(k_1 \cdot k_2)+(1-\epsilon)k_1^\mu k_2^\nu+(1-\epsilon)k_2^\mu k_1^\nu\right) \,, \\
  \widetilde{\cal S}(k_1,k_2)=&-\frac{(p_1 \cdot p_2)^2}{2(p_1 \cdot k) (p_2 \cdot k)}\left(\frac 2{(p_1 \cdot k_1)(p_2 \cdot k_1)}+\frac 1{(p_1 \cdot k_1)(p_2 \cdot k_2)}\right)
   +\frac {(p_1 \cdot p_2)}{k^2}\frac{2}{(p_1 \cdot k_1)(p_2 \cdot k_2)}
 \nn\\[0.5ex]&
 -\frac{(p_1 \cdot p_2)}{2 k^2 (p_1 \cdot k)(p_2 \cdot k)}\frac{((p_1 \cdot k_1)(p_2 \cdot k_2)-(p_1 \cdot k_2)(p_2 \cdot k_1))^2}{(p_1 \cdot k_1)(p_2 \cdot k_2)(p_1 \cdot k_2)(p_2 \cdot k_1)}
+(k_1\leftrightarrow k_2)\,,
\end{align}
with $k=k_1+k_2$.

While the phase space integration for the single-gluon emission at one loop is elementary, similar to the case of $w_{i,{\rm LO}}$, in the contributions arising from the emission of two soft partons we need to carry out an additional integration over $k_1$ and $k_2$ at fixed $k=k_1+k_2$.
These integrals have been computed in Ref.~\cite{Catani:2023tby} in the context of the evaluation of the soft contributions to heavy-quark production.
In the following we highlight the main steps of their computation.

We start from the $q\bar q$ contribution.
Since we want to integrate at fixed $k=k_1+k_2$, the factors $J_\mu(k_1+k_2)$ 
can be trivially integrated by using the delta function 
\begin{align}
&\int
  \frac{d^d k_1}{(2\pi)^{d-1}} \delta_+(k_1^2)
  \frac{d^d k_2}{(2\pi)^{d-1}} \delta_+(k_2^2)
    |J_i^{q\bar q}(k_1,k_2)|^{2}_{(\text{LO})}
  \delta^{(d)}\left(k- k_1 -k_2\right)=
  \\ \nn
  &\hspace{9.cm}
 =(4\pi\, \alpha_S(\mu_R^2) \, \mu_R^{2\ep} S_\ep^{-1})^2 \,C_i\, \frac{{\widehat J}_\mu(k){\widehat J}_\nu(k)}{(2\pi)^{d-1}}F^{\mu\nu}(k)
\end{align}
and we only need to compute the following quantity
\beq
F^{\mu\nu}(k) =
  \int
  \frac{d^d k_1\,d^d k_2}{(2\pi)^{d-1}} \delta_+(k_1^2) \delta_+(k_2^2)
  \Pi^{\mu\nu}(k_1,k_2) \delta^{(d)}\left(k- k_1 -k_2\right)\;.
\eeq

The tensor $F^{\mu\nu}$ depends only on $k$, and in addition satisfies $k_\mu k_\nu F^{\mu\nu} = 0$. This implies that its structure is $F^{\mu\nu}(k) = C (-g^{\mu\nu}+k^\mu k^\nu/k^2)$, with $C$ being a constant to be determined. We can extract the value of $C$, for instance, from the computation of $g_{\mu\nu}F^{\mu\nu}(k)$. We find
\beq
C = \frac{T_R  \,  (1-\ep) (k^2)^{-1-\ep}}{\Gamma(\tfrac{5}{2}-\ep) 16^{1-\ep}\,\pi^{\tfrac32 -\ep} } \;.
\eeq

We now consider the integration of the double-gluon current.
The procedure to compute the integral of the first term is identical to the one we just described, the only difference being the structure of the tensor $\widetilde \Pi^{\mu\nu}$ with respect to $\Pi^{\mu\nu}$. By following the same steps, it can be proven that its contribution to $w_{i,\text{NLO}}$ can be simply obtained by replacing
\beq
\label{eq:repl}
n_F T_R \, \to - C_A \frac{11-7\ep}{4(1-\ep)} 
\eeq
in the $q \bar q$ result. The replacement in Eq.~(\ref{eq:repl}) also includes the symmetry factor $1/2$ in Eq.~(\ref{eq:web}), due to the emission of two identical gluons.

The last remaining contribution in the double-gluon soft current is the one proportional to $\widetilde{\cal S}(k_1,k_2)$. We proceed by integrating over $k_1$ against the momentum conservation delta function. Then, we consider the rest frame of $k$ and integrate the energy component and the modulus of the $(d-1)$ spacial components against the two remaining delta functions. In this way, we are left only with angular integrals.
By following the procedure described above, we arrive to the intermediate expression, 
\begin{align}
&\int
\frac{d^dk_1}{(2\pi)^{d-1}} \, \delta_+(k_1^2)\, \frac{d^dk_2}{(2\pi)^{d-1}}\,\delta_+(k_2^2) \,\delta^{(d)}(k-k_1-k_2)
\,\widetilde{\cal S}(k_1,k_2)\;= 
\\
& \;\;\;\;\;\;\;
=\frac{\pi^{-11/2+3\ep}}{64^{1-\ep}\Gamma(\tfrac{1}{2}-\ep)}\frac{(k^2)^{-1-\ep}(p_1 \cdot p_2)}{(p_1 \cdot k)(p_2 \cdot k)}\nn
\left[\left(2 - \frac{k^2 (p_1 \cdot p_2)}{(p_1 \cdot k)(p_2 \cdot k)}\right)\AngInt_{11}^+-2
\frac{k^2 (p_1 \cdot p_2)}{(p_1 \cdot k)(p_2 \cdot k)}
  \AngInt_{11}^-+2\AngInt_{10}\right] \;,
\end{align}
where the angular integrals $\AngInt_{ij}^\pm$ are defined as

 \begin{equation}\label{eq:angular}
 \AngInt_{i,j}^\pm=\int_0^\pi d\theta \int_0^\pi d\phi \, \frac{\sin^{d-3}\theta \sin^{d-4}\phi}
{(1- \cos\theta)^i(1\pm  \cos\chi \cos\theta\pm  \sin\chi \sin\theta \cos\phi)^j}\;,
 \end{equation}
with
\begin{equation}
\cos\chi =
1-\frac{k^2 (p_1 \cdot p_2)}{(p_1 \cdot k)(p_2 \cdot k)}\,
.
\end{equation}
The results for the required angular integrals can be found for instance in Ref.~\cite{Somogyi:2011ir}.

After putting together all the pieces whose calculation was described above, we arrive to the following $d$-dimensional result: \begin{align}\label{eq:web_NLO}
  w_{i,{\rm NLO}}&(k;\ep) = \; C_i \left(\frac{\mu_R^2}{k_T^2}\right)^{2\ep} \frac{c(2\ep)}{k_T^4} \left(\frac{\as(\mu_R^2)}{\pi}\right)^2  \\
&  \times \Bigg\{
  \left[
    -\left(\frac{\mu_R^2}{k_T^2}\right)^{-\ep }\frac{c(\ep )}{c(2\ep )}\frac{ 
      \left(11 C_A-4 T_R n_F\right)}{6 \ep }- C_A \frac{\pi ^2 \cos (\pi
       \ep )}{ \sin^2 (\pi  \ep )}
  \right]
  \delta(1-t) \nonumber\\
    & - \left((12-9\ep) C_A-4T_R
    n_F\ep\right)\frac{(1-\ep)^2
     \Gamma (1-2 \ep ) }{\ep  \Gamma (4-2 \ep )}(1-t)^{-1-\ep} t^{2+\ep}+C_A \frac{2 \pi}{\sin (\pi  \ep )}(1-t)^{-1-2\ep} t^{2+2\ep} \nonumber\\
  & + C_A  \left[\frac{2}{1+\ep} \, _2F_1(1,1;2+\ep;1-t) -\frac{1}{1-\ep} \,
    _2F_1(1,1;2-\ep ;1-t) \right] (1-t)^{-\ep} t^{2+\ep} \nn
  \Bigg\}\;,
\end{align}
where we have introduced the dimensionless variable $t$, which is defined as
\begin{equation}
    t=k_T^2/m_T^2\, ,~~~~~~~~~0\leq t \leq 1\, .
\end{equation}
The expression in Eq.~(\ref{eq:web_NLO}) is apparently singular: there are explicit poles of the dimensional regularisation parameter $\ep$, plus there are endpoint divergences at $t=1$ for which keeping the expression in $d$-dimensions seems necessary for their regularisation.
However, this is not the case, and we can make a more direct connection to the $d=4$ result by expanding the factors $(1-t)^{-1-a \ep}$ in terms of plus distributions and rearranging the different terms. We arrive to the following result
\beq
\label{wnlotot}
w_{i,\text{NLO}}(k;\ep)=
C_i \left(\frac{\mu_R^2}{k_T^2}\right)^{2\ep} \frac{c(2\ep)}{k_T^4} \left(\frac{\as(\mu_R^2)}{\pi}\right)^2
t^{2+\ep} (w_{\text{NLO}}^{a}+w_{\text{NLO}}^{b})\,,
\eeq
where $w_{\text{NLO}}^{a}$ and $w_{\text{NLO}}^{b}$ are given by
\beeq
\label{wnloa}
w_{\text{NLO}}^{a}&=&
\frac{67 C_A-20 T_R n_F-\left(44 C_A-16 T_R n_F\right)\epsilon}{6 (3-2\epsilon) (1-2\epsilon)}\,\delta (1-t) 
\nn\\
&-&
\frac{11 C_A-4 T_R n_F-\left(7 C_A-4 T_R n_F\right)\epsilon}{6 - 8 (2-\epsilon) \epsilon
}\,[(1-t)^{-1-\epsilon}]_+
    \nn\\
   &+&C_A (1-t)^{-\epsilon }\left(
\frac{2}{1+\ep} \, _2F_1(1,1;2+\ep;1-t) -\frac{1}{1-\ep} \,
    _2F_1(1,1;2-\ep ;1-t)
   \right)\,, \\
   \label{wnlob}
    w_{\text{NLO}}^{b}&=&\delta (1-t) \left(\frac{\left(11 C_A-4 T_R n_F\right)}{6\epsilon }\left(1-\frac{c(\epsilon )}{c(2 \epsilon )}\left(\frac{\mu^2}{k_T^2}\right)^{-\epsilon }\right)+\epsilon ^2 C_A
   f(\epsilon )\right)\nonumber\\
   &+& 2 C_A\,\frac{1}{\epsilon}\,\left(\frac{\pi  \epsilon}{\sin (\pi  \epsilon )}\, t^{\epsilon } [(1-t)^{-1 -2\epsilon}]_+-[(1-t)^{-1-\epsilon}]_+\right)\,,
\eeeq
with $f$ being a function of $\epsilon$ defined as
\beq
f(\epsilon)= \frac 1{\epsilon^4}\,\left(2-\frac{\pi\,\epsilon}{\sin(\pi\,\epsilon)}-\cos(\pi\,\epsilon)\frac{(\pi\,\epsilon)^2}{\sin^2(\pi\,\epsilon)}\right)
= \frac{7\pi^4}{180} + {\cal O}(\ep^2)\,.
\eeq
In order to take the $d\to 4$ limit, we can directly replace $\ep=0$ in $w_{\text{NLO}}^{a}$.
The expression of $w_{\text{NLO}}^{b}$ still has explicit poles in $\ep$, though it is straightforward to show upon expansion that it is also finite in the $\ep\to 0$ limit.

Using our result in Eq.~(\ref{wnlotot})
for $w_{i,{\rm NLO}}(k;\ep)$, we can compute the soft-gluon effective coupling at order $\as^2$.
We consider the generalised definition in Eq.~(\ref{eq:softcoupling_f}).
It is possible to obtain compact results for the functional form 
$\cf(k_T^2/m_T^2)=(k_T/m_T)^{2p}$,
which corresponds to the definition of $\ATt$ and $\Aot$ for $p=0$ and $p=1$, respectively, and it smoothly interpolates between them for $p \in (0,1)$.
The integrals over $k_T^2$ and $m_T^2$ in Eq.~(\ref{eq:softcoupling_f}) can be performed in a relatively straightforward way (one of them is trivial due to the presence of the delta function). The most complicated terms are those containing hypergeometric functions in Eq.~(\ref{wnloa}), and the computation is more easily carried out by using an integral representation of ${}_2F_1$ and exchanging the order of integration, thus leading to hypergeometric functions of the type 
${}_3F_2$.
We obtain the following result:
\begin{eqnarray}\label{eq:A2tp}
\widetilde{\mathscr{A}}^{(2)}_{\cf(t)=t^p,i}(\ep) &=&
C_i\,\bigg\{
-\frac{c(\ep)(11C_A-4 T_R n_F)}{12\ep}
+ \frac{c(2\ep)\,g(\ep,p)}{\ep} \frac{[C_A(11-7\ep)-4 T_R n_F(1-\ep)]}{4(3-2\ep)(1-2\ep)}
\\&+& \frac{C_A \, c(2\ep) \, g(\ep,p) \, r_p(\ep)}{2(1-2p\ep)} 
- \frac{C_A \, c(2\ep)}{2\ep^2}
\left[
\frac{(\pi\ep)^2 \cos(\pi\ep)}{\sin^2(\pi\ep)}  +\frac{(\pi\ep) \, g(2\ep,p/2)}{\sin(\pi\ep)}  -2g(\ep,p)
\right]
\bigg\} \nonumber ,
\end{eqnarray}
with
\begin{eqnarray}
r_p(\ep) &=& \frac{2}{1+\ep}\, {}_3F_2(1,1,1-\ep;2-2p\ep,2+\ep;1)
-\frac{1}{1-\ep}\, {}_3F_2(1,1,1-\ep;2-2p\ep,2-\ep;1)\,, \\
g(\ep,p)&=&\frac{\Gamma(1-\ep) \, \Gamma(1+\ep-2p\ep)}{\Gamma(1-2p\ep)}\, .
\end{eqnarray}
Its $\ep$ expansion up to ${\cal O}(\ep^2)$ is
\begin{align}\label{eq:A2tpExp}
\widetilde{\mathscr{A}}^{(2)}_{\cf(t)=t^p,i}(\ep) &= A_i^{(2)} +\ep\, C_i \left[C_A \left( \frac{101}{27} - \frac{11\, \pi^2}{144}(1+4p) -\frac{7 \zeta_3}{2}\right) 
+  2 T_R n_F \left(-\frac{14}{27} + \frac{\pi^2}{72}(1+4p) \right) \right] 
  \nn\\
&+\ep^2\, C_i \bigg[ C_A \left( \frac{607}{81} - \frac{67\, \pi^2}{216}(1+2p) -\frac{11 \zeta_3}{36}(7-6 p+12 p^2) - \frac{\pi^4}{360}(21-31 p+20 p^2)\right) \nn \\
&+ 2 T_R n_F \left(-\frac{82}{81} + \frac{5\,\pi^2}{108}(1+2p) + \frac{\zeta_3}{18}(7-6 p+12 p^2) \right) \bigg] +{\cal{O}}(\ep^3) \;\;.
\end{align}
The results in Eqs.~(\ref{eq:A2tp}) and (\ref{eq:A2tpExp}) for the particular cases $p=0$ and $p=1$ were anticipated in Ref.~\cite{Catani:2019rvy}.

\section{Properties of the generalised soft effective coupling}
\label{s:general}

In this section we discuss some general features of the soft-gluon effective coupling.

We consider the generalised effective coupling in Eq.~(\ref{eq:softcoupling_f}).
The scale $\mu$ of the effective coupling is specified through the function
$\cf(t)$ with $t=k_T^2/m_T^2$. The function $\cf(t)$ is a smooth function of $t$ in the interval $0 \leq t \leq 1$, with the following behaviour at the endpoints $t=0,1$.
In the limit $t \to 0$ 
we have $t \ln^n \cf(t) \to 0$ for any positive integer $n=1,2,\dots$, so that the integrand
in the right-hand side of Eq.~(\ref{eq:softcoupling_f}) is integrable in the highly off-shell region where $m_T \gg k_T$. In the limit $t \to 1$ (i.e., in the on-shell limit
$k^2 \to 0$) we have $\cf(t) \to \cf(1)$, where $\cf(1)$ is a finite (non-vanishing) constant value. The finite value of $\cf(1)$ is a consequence of the fact that in the on-shell region we have $k_T=m_T$ and, therefore, in this region there is basically only a single scale (modulo its overall normalisation) that can be used to define the effective coupling through
the integration of the probability density $w_i(k;\ep)$. Actually, even the overall normalisation of the scale has a trivial effect on the effective coupling, since from 
Eq.~(\ref{eq:softcoupling_f}) we have
\beq
\label{af1}
\Aft (\as(\mu^2);\ep) = \Afft(\as(\mu^2\cf(1));\ep) \;\;.
\eeq
Therefore, as already mentioned below Eq.~(\ref{eq:softcoupling_f}), in the following
we limit ourselves to considering the case with $\cf(1)=1$
(the more general case with $\cf(1)\neq 1$ can be simply recovered through the replacements
$\cf(t) \to \cf(t)/\cf(1)$, $\mu^2 \to \mu^2\cf(1)$).

The generalised soft effective coupling is obtained through Eq.~(\ref{eq:softcoupling_f})
by integrating the probability density $w_i(k;\ep)$. We introduce the rescaled density
${\widehat w}_i(k;\ep)$, which is defined as follows
\beq
\label{what}
{\widehat w}_i(k;\ep) \equiv \frac{1}{2} \;m_T^4 \;w_i(k;\ep) \;\;.
\eeq
The function ${\widehat w}_i(k;\ep)$ is dimensionless, and it depends on the renormalisation scale $\mu_R$ of $\as$ and on the scales $k_T$ and $m_T$. We find it
convenient to specify the perturbative expansion of ${\widehat w}_i(k;\ep)$ in the following form:
\beq
\label{wtexp}
{\widehat w}_i(k;\ep) = \ATt(\as(k_T^2);\ep) \;\delta(1-t)
+ \sum_{n=2}^{+\infty} \left( \frac{\as(k_T^2)}{\pi} \right)^n \;
\big[ {\widehat w}_{T,i}^{(n)}(t;\ep) \big]_+
 \;\;,
\eeq
where $\ATt$ is the effective coupling in Eq.~(\ref{eq:defAT}),
and $\big[ {\widehat w}^{(n)}(t;\ep) \big]_+$ is the plus-distribution over $t$ 
of the function ${\widehat w}^{(n)}(t;\ep)$.

The structure of Eq.~(\ref{wtexp}) can be easily explained.
To introduce the perturbative expansion in Eq.~(\ref{wtexp}) we have first set $\mu_R=k_T$
in the scale of $\as$. Then the corresponding perturbative coefficients are dimensionless
and depend on the variables $t$ and $\ep$. As we know from the explicit LO and NLO computations of ${w}_i(k;\ep)$ (see Sects.~\ref{sec:def} and \ref{sec:pert}), the $t$ dependence of these perturbative coefficients includes singular distributions such as 
$\delta(1-t)$ and plus-distributions of singular functions at $t=1$ (e.g., 
$[(1-t)^{-1} \ln^k(1-t)]_+$). In Eq.~(\ref{wtexp}) the $t$ dependence is entirely expressed 
in terms of $\delta(1-t)$ and plus-distributions of {\it generic} functions 
${\widehat w}_{T,i}^{(n)}(t;\ep)$, which are not necessarily singular functions of $t$ in the limit $t \to 1$. We note that the series in the right-hand side of Eq.~(\ref{wtexp})
has $n \geq 2$, since the LO term with $n=1$ is fully proportional at $\delta(1-t)$.
Inserting Eqs.~(\ref{what}) and (\ref{wtexp}) in Eq.~(\ref{eq:defAT}), we note that the terms $\big[ {\widehat w}^{(n)}(t;\ep) \big]_+$ give vanishing contributions to the integration over $m_T$ and, consequently, the factor in front of $\delta(1-t)$ 
in Eq.~(\ref{wtexp}) is (by definition) the effective coupling $\ATt$ of 
Eq.~(\ref{eq:defAT}).
 
The explicit expression of the NLO function ${\widehat w}_{T,i}^{(2)}(t;\ep)$ in 
Eq.~(\ref{wtexp}) directly follows from Eqs.~(\ref{wnlotot})--(\ref{wnlob}), and 
we have
\beeq
\label{wh2}
{\widehat w}_{T,i}^{(2)}(t;\ep)
&=& \frac{C_i}{2} \;c(2\ep) \;t^{\ep} \Bigl\{
 - \frac{11 C_A-4T_R n_F-\left(7 C_A-4T_R n_F\right)\epsilon}{6 - 8 (2-\epsilon) \epsilon
}\,(1-t)^{-1-\epsilon}
\Bigr.    
\nn\\
   &+&C_A (1-t)^{-\epsilon }\left(
\frac{2}{1+\ep} \, _2F_1(1,1;2+\ep;1-t) -\frac{1}{1-\ep} \,
    _2F_1(1,1;2-\ep ;1-t)
   \right) 
\nn \\
&+& \Bigl.
2 C_A\,\frac{1}{\epsilon}\,(1-t)^{-1-\epsilon }\,
\left(\frac{\pi  \epsilon}{\sin (\pi  \epsilon )}\, t^{\epsilon } (1-t)^{-\epsilon}-
1\right) \Bigr\} \;\;.
\eeeq
In particular, in the four-dimensional case we have
\beq
\label{wh20}
{\widehat w}_{T,i}^{(2)}(t;\ep=0) = \frac{C_i}{2(1-t)} \left[
-\frac{1}{6}(11 C_A-4T_R n_F) + C_A \ln\frac{t}{(1-t)^2}
\right] \;\;.
\eeq

The perturbative representation in Eq.~(\ref{wtexp}) can be used to obtain a general master formula that relates soft effective couplings that are specified by different scale functions $\cf(t)$. Inserting Eqs.~(\ref{what}) and (\ref{wtexp}) in 
Eq.~(\ref{eq:softcoupling_f}), we straightforwardly obtain
\beq
\label{mastert}
\Aft(\as(\mu^2);\ep) - \ATt(\as(\mu^2);\ep) = \sum_{n=2}^{+\infty} 
\frac{1}{\pi^n} \int_0^1 dt \,\bigl[ \as^n(\mu^2 \cf(t)) - \as^n(\mu^2) \bigr] \,
\; {\widehat w}_{T,i}^{(n)}(t;\ep) \;\;.
\eeq
Then, considering Eq.~(\ref{mastert}) for two different functions $\cf_1(t)$ and
$\cf_2(t)$, we equivalently obtain
\beq
\label{masterf}
\Afto(\as(\mu^2);\ep) - \Aftt(\as(\mu^2);\ep) = \sum_{n=2}^{+\infty} 
\frac{1}{\pi^n} \int_0^1 dt \,\bigl[ \as^n(\mu^2 \cf_1(t)) - \as^n(\mu^2\cf_2(t)) \bigr] \,
\; {\widehat w}_{T,i}^{(n)}(t;\ep) \;\;.
\eeq
We note that the coupling $\as(k_T^2)$ in Eq.~(\ref{wtexp}) becomes the $t$-dependent function $\as(\mu^2 \cf(t))$ after its insertion in Eq.~(\ref{eq:softcoupling_f}).
Then, the action of the plus-distribution of ${\widehat w}_{T,i}^{(n)}(t;\ep)$ leads to the subtraction term $[ \as^n(\mu^2 \cf(t)) - \as^n(\mu^2\cf(1)) ]$ (we recall that we set 
$\cf(1)=1$) in the right-hand side of Eq.~(\ref{mastert}).

The master formula in Eq.~(\ref{mastert}) (or, equivalently, Eq.~(\ref{masterf}))
gives a relation between the soft effective couplings. In particular,
the difference between the soft effective couplings is directly controlled by the difference
of the QCD coupling $\as$ at the scales that specify the effective couplings.
This implies that the difference in the soft effective couplings has basically an UV origin
from the running behaviour of 
$\as$. This feature is somehow expected since the soft effective coupling is IR/collinear safe and dimensionless and, therefore,
the dependence on the corresponding scale momentum function $\cf(k_T^2/m_T^2)$
can only occur through the scaling violation due to the scale dependence of 
$\as$.

Using Eq.~(\ref{mastert}) or (\ref{masterf}), we can derive important relations between the soft effective couplings, the cusp anomalous dimension and the CMW coupling. We discuss
these relations in turn.

The cusp anomalous dimension is a relevant quantity that controls the evolution of the parton distribution functions in the soft limit  \cite{Korchemsky:1988si, Moch:2004pa, Vogt:2004mw},
the renormalisation of cusped 
light-like Wilson line operators \cite{Korchemsky:1987wg}
and the IR divergences of QCD scattering amplitudes
\cite{Catani:1998bh, Aybat:2006mz, Gardi:2009qi, Becher:2009qa, Almelid:2015jia}.
The cusp anomalous dimension $A_i(\as)$ can be expressed as a power series expansion in $\as$ in the following form:
\beq
\label{cuspexp}
A_i(\as) = \sum^{+\infty}_{n=1} \left(\frac{\as}{\pi}\right)^n A_i^{(n)} \;\;,
\eeq
and it is presently known up to ${\cal O}(\as^4)$ 
\cite{Henn:2019swt, vonManteuffel:2020vjv}.
We recall that up to ${\cal O}(\as^2)$ the cusp anomalous dimension is equal to the CMW
coupling \cite{Catani:1990rr}. Beyond ${\cal O}(\as^2)$, $A_i(\as)$ is not directly related to physical observables, since it refers to IR or UV divergent quantities and, in particular, it is related to $\msbar$-scheme factorisation of IR and UV divergences in the context of dimensional regularisation in $d=4 - 2\ep$ space-time dimensions.

Using Eq.~(\ref{mastert}) and the results of Ref.~\cite{Catani:2019rvy}, we obtain the following all-order relation between the cusp anomalous dimension and the $d$-dimensional soft effective coupling:
\beq
\label{cuspid}
\Aft(\as; \,\ep=\beta(\as)) = A_i(\as) \;\;.
\eeq
In this relation the $d$-dimensional soft effective coupling is evaluated at the conformal
point $\ep=\beta(\as)$, where the $d$-dimensional QCD $\beta$-function 
$\beta(\as) - \ep$ vanishes. The relation in Eq.~(\ref{cuspid}) implies that
all soft effective couplings are equal at the conformal point $\ep=\beta(\as)$ and, moreover, they
are equal to the cusp anomalous dimension. The relation (\ref{cuspid}) is not specific of QCD, and it also applies to other gauge theories. In particular, in the case of
${\cal N}=4$ maximally supersymmetric Yang--Mills theory we have $\beta(\as)=0$ and,
therefore, the cusp anomalous dimension coincides with the physical (four-dimensional)
soft effective coupling: ${\mathscr A }_{\cf}(\as) = A(\as)$.

The equality between $A_i(\as)$ and the two effective couplings $\ATt$ and $\Aot$
was first pointed out in Ref.~\cite{Catani:2019rvy}. Therefore Eq.~(\ref{cuspid})
generalises this result by adding that all the soft effective couplings are equal at the conformal point. The conformal-point equality of the soft effective couplings is a straightforward
consequence of Eq.~(\ref{masterf}). Indeed, setting $\ep=\beta(\as)$, the $d$-dimensional
QCD coupling $\as(\mu^2)$ does not depend on the scale $\mu$ and, consequently,
the difference $[ \as^n(\mu^2 \cf_1(t)) - \as^n(\mu^2\cf_2(t)) ]$ in the right-hand side of
Eq.~(\ref{masterf}) vanishes.

In the right-hand side of Eqs.~(\ref{mastert}) or (\ref{masterf}) we can express the
$d$-dimensional QCD coupling $\as(\mu^2 \cf)$ in terms of $\as(\mu^2)$ and $\cf$, and we can obtain the difference between the soft effective couplings order-by-order in perturbation theory.
 
At ${\cal O}(\as^2)$ we find the following relation between the perturbative coefficients
$\Aft^{(2)}$ of the effective couplings:
\beeq
\label{at2dif}
\Afto^{(2)}(\ep) - \Aftt^{(2)}(\ep) &=&
 \int_0^1 dt \,\bigl[ \left(\cf_1(t)\right)^{-2\ep} - \left(\cf_2(t)\right)^{-2\ep} \bigr] \,\; {\widehat w}_{T,i}^{(2)}(t;\ep) \nn \\
&=& 2 \,\ep \int_0^1 dt \,\ln\left[\frac{\cf_2(t)}{\cf_1(t)}\right] 
\;{\widehat w}_{T,i}^{(2)}(t;\ep=0) + {\cal O}(\ep^2)  \;\;,
\eeeq
where ${\widehat w}_{T,i}^{(2)}$ is explicitly given in Eqs.~(\ref{wh2}) and (\ref{wh20}).
We note that the difference in Eq.~(\ref{at2dif}) is of ${\cal O}(\ep)$ and, therefore,
all the physical (four-dimensional) soft effective couplings are equal up to 
${\cal O}(\as^2)$ (the equality at ${\cal O}(\as)$ follows from Eq.~(\ref{a1coeff})),
as anticipated in Eqs.~(\ref{af2}) and (\ref{a2coef}). We recall that 
$\AT^{(2)}$ \cite{Banfi:2018mcq, Catani:2019rvy} and $\Ao^{(2)}$ \cite{Catani:2019rvy} 
are equal to the second-order coefficient of the CMW coupling and, consequently,
the remarkable $\cf$ independence of $\Af^{(2)}$ is consistent with the universality of the CMW coupling \cite{Catani:1990rr}.

Expanding Eq.~(\ref{masterf}) up to ${\cal O}(\as^3)$  we can directly relate the perturbative coefficients $\Af^{(3)}$ of the four-dimensional soft effective couplings.
We find
\beq
\label{a3dif}
\Afone^{(3)} - \Aftwo^{(3)} = 2 \pi \beta_0 \, \int_0^1 dt \,\ln\left[\frac{\cf_2(t)}{\cf_1(t)}\right] 
\;{\widehat w}_{T,i}^{(2)}(t;\ep=0)  \;\;.
\eeq  
We note that the difference between the effective couplings $\Af$ at ${\cal O}(\as^3)$
is proportional to the first-order coefficient $\beta_0$ of the QCD $\beta$-function. 
In particular, Eq.~(\ref{a3dif}) shows how this third-order difference can be computed
from the knowledge of the probability density $w(k;\ep)$ at ${\cal O}(\as^2)$
(i.e., without knowing $w(k;\ep)$ at ${\cal O}(\as^3)$). 
The third-order coefficients $\AT^{(3)}$ and $\Ao^{(3)}$ are known from 
Refs.~\cite{Banfi:2018mcq, Catani:2019rvy} and \cite{Catani:2019rvy}, respectively. 
Using these third-order results and the simple $t$-integral in Eq.~(\ref{a3dif}),
one can straightforwardly evaluate $\Af^{(3)}$ for various scale momentum
functions $\cf(t)$.
For instance, considering the function $\cf(t)=t^p$ (as in Sect.~\ref{sec:pert}) and the expression of
${\widehat w}_{T,i}^{(2)}$ in Eq.~(\ref{wh20}), we obtain
\beq
\label{a3ft}
\Afp^{(3)} - \AT^{(3)} = - 2 \pi \beta_0 \,p\, \int_0^1 dt \,\ln t 
\;\,{\widehat w}_{T,i}^{(2)}(t;\ep=0)  
= - \frac{\pi^2}{3} \,p \left(\pi \beta_0\right)^2 C_i\;\;.
\eeq
Setting $p=1$ in Eq.~(\ref{a3ft}) we recover the difference $(\Ao^{(3)} - \AT^{(3)})$
that was first computed in Ref.~\cite{Catani:2019rvy}. In the case of generic values of $p$,
Eq.~(\ref{a3ft}) is a new result. We note that the difference in Eq.~(\ref{a3ft}) is proportional to $\beta_0^2$ (rather than simply to $\beta_0$ as in Eq.~(\ref{a3dif})):
such proportionality is `accidental' since it is due to the fact that the integral
$\int_0^1 dt \,\ln t \;{\widehat w}_{T,i}^{(2)}(t;\ep=0)$ turns out to be proportional to
$\beta_0$.

In four space-time dimensions the difference
$[ \as^n(\mu^2 \cf) - \as^n(\mu^2) ]$ is proportional to the coefficients of $\beta(\as)$.
Therefore, Eq.~(\ref{masterf}) shows that, in general, the difference 
$\Afone^{(n)} - \Aftwo^{(n)}$ with $n \geq 4$ is proportional to the coefficients
$\beta_0, \beta_1, \cdots, \beta_{n-3}$. 

\section{All-order expressions in the 
large-\texorpdfstring{$\mathbf{n_F}$}{nF} limit}

\label{sec:largenf}

In this section, we present results for the probability density 
$w_i$
and the soft coupling $\Aft$
in the large-$n_F$ limit $n_F\gg 1$. In this framework, it is possible to obtain compact results at any order $n$ in the $\as$ expansion and to sum them to get expressions valid at all orders.

Only two different structures contribute to this limit at fixed perturbative order: 
those coming from the renormalisation of $\as$
and 
those arising from multiple quark-bubble insertions to the quark-antiquark soft current.
Within the renormalisation procedure of $\as$
only the contributions that depend on $\beta_0$ are relevant in the large-$n_F$ limit.
The resulting series of terms with an increasing number of powers of $\beta_0$ can be summed at all orders as a geometric series, leading to a factor $(1+\beta_0^{(n_F)}\as/\ep)^{-1}$, where $\beta_0^{(n_F)}=-T_R n_F/(3\pi)$ is the
$n_F$-dependent part of the coefficient $\beta_0$.
The contribution coming from the sum of bubble insertions into the gluon propagator is also a geometric series at the amplitude level, since only quark loops contribute in this limit.
 We find
\begin{align}\label{eq:web_large_nf_old}
\sum_{n=0}^\infty w_{i,\text{N$^n$LO}}(k;&\ep)\Big\vert_{n_F\gg 1} =
\frac{2C_i c(\ep)}{k_T^4} \frac{\as}{\pi} 
  \left(
  1+\frac{\beta_0^{(n_F)}}{\ep}\as
  \right)^{-1} \left(\frac{\mu_R^2}{k_T^2}\right)^\ep
\Bigg\{
 \delta(1-t) 
\\ &-
\frac{\sin(\pi\ep)}{\pi}h(\as,\ep) \left(\frac{\mu_R^2}{k_T^2}\right)^{\ep}t\left(\frac{1-t}t\right)^{-1-\ep} \left|1-h(\as,\ep)\, e^{i \pi \ep}\left(\frac{\mu_R^2}{k_T^2}\right)^{\ep}\left(\frac{1-t}t\right)^{-\ep}\right|^{-2}\Bigg\}\nn\;,
\end{align}
where we have defined:
\beq
h(\as,\ep) =  -\frac{\Gamma^2(2-\ep)}{\Gamma(4-2\ep)}\frac{\pi}{\sin(\pi\ep)}
\frac{\as}{\pi}
  \left(
  1+\frac{\beta_0^{(n_F)}}{\ep}\as
  \right)^{-1} c(\ep) \, 2 T_R n_F \,.
\eeq
The last factor in the second term of Eq.~(\ref{eq:web_large_nf_old}) comes from squaring the (complex) geometric sum that contributes at the amplitude level. It can be rewritten in a more convenient way as follows
\beeq \label{eq:web_large_nf}
\sum_{n=0}^\infty w_{i,\text{N$^n$LO}}(k;\ep)\Big\vert_{n_F\gg 1} &=&
\frac{2C_i c(\ep)}{k_T^4} \frac{\as}{\pi}
  \left(
  1+\frac{\beta_0^{(n_F)}}{\ep}\as
  \right)^{-1} \left(\frac{\mu_R^2}{k_T^2}\right)^\ep
\Bigg\{
   \delta(1-t) 
\\ &-&
\frac{t^2}{1-t} \,\frac{1}{\pi} \,\text{Im}\left[\left(1-h(\as,\ep)\, e^{ i \pi \ep}\left(\frac{\mu_R^2}{k_T^2}\right)^{\ep}\left(\frac{1-t}t\right)^{-\ep}\right)^{-1}\right]\Bigg\}\,.\nn
\eeeq
We note that in Eqs.~(\ref{eq:web_large_nf_old}) and (\ref{eq:web_large_nf})
for the probability density $w_i$, the strong coupling $\as$ is always evaluated at the arbitrary renormalisation scale $\mu_R$. For compactness, this explicit dependence is omitted
and we simply write $\as(\mu_R^2) = \as$.

We now compute the soft-gluon effective coupling at large $n_F$, based on the result in Eq.~(\ref{eq:web_large_nf}). We start by considering a generic function $\cf(t)$ and we apply the definition of the soft coupling (Eq.~(\ref{eq:softcoupling_f})) to each term of the geometric series present in the second term of Eq.~(\ref{eq:web_large_nf}). After some simplification and explicitly taking the imaginary part, we arrive to the following result:
\begin{eqnarray}\label{eq:large_nf_general_f}
\sum_{n \geq 1}
\left[
\left(
\frac{\as}{\pi}
\right)^n
\Aft^{(n)}(\ep)
\Big\vert_{n_F\gg 1}
\right]
&=&
C_i \, c(\ep) \, \frac{\as/\pi}{1-\frac{1}{\ep}\frac{\as}{\pi}T_R \frac{n_F}{3}}\,
\Bigg\{ 1 +
\sum_{m\geq 1} \Bigg(
\frac{-\sin(m \pi \ep)}{\pi}
 \\
 &\times&
 \left[
h(\as,\ep)
\right]^m
\int_0^1 dt\,
\left[\cf(t)\right]^{-(1+m)\,\ep}\, t^{m\,\ep}\,(1-t)^{-1-m\,\ep}
\Bigg)
\Bigg\} . \nonumber
\end{eqnarray}
where the first term in the curly bracket arises from the $\delta(1-t)$ term in Eq.~(\ref{eq:web_large_nf}) and we have used $\cf(1) = 1$, while the remaining terms in the sum are the ones coming from the second term in the bracket of Eq.~(\ref{eq:web_large_nf}).
In the equation above, as well as in the following results for the soft-gluon effective coupling, $\as$ is evaluated at the scale $\mu$ given by $\mu^2=k_T^2/\cf(k_T^2/m_T^2)$. This dependence is left implicit for compactness,
and we simply write $\as(\mu^2) = \as$.

Considering explicit forms for the function 
$\cf(t)$, the expression in Eq.~(\ref{eq:large_nf_general_f}) can be further
worked out.
For the particular case $\cf(t)=t^p$, the integral over $t$ can be performed in a closed form and we find
\begin{equation}\label{eq:large_nf_tp}
\sum_{n \geq 1}
\left[
\left(
\frac{\as}{\pi}
\right)^n
{\widetilde\mathscr{A}}_{\cf(t)=t^p,i}^{(n)}(\ep)
\Big\vert_{n_F\gg 1}
\right]
=
C_i \, c(\ep) \, \frac{\as/\pi}{1-\frac{1}{\ep}\frac{\as}{\pi}T_R \frac{n_F}{3}}\,
\sum_{m \geq 0}
\left[
h(\as,\ep)
\right]^m
\frac{\Gamma(1+ m \ep(1-p)-p \ep)}{\Gamma(1+ m \ep) \Gamma(1 -p (1+m) \ep)} \,.
\end{equation}

In the $p=0$ case the series in Eq.~(\ref{eq:large_nf_tp}) can be explicitly summed to obtain
\begin{equation}\label{eq:ATallorders}
\sum_{n \geq 1}
\left[
\left(
\frac{\as}{\pi}
\right)^n
{\widetilde\mathscr{A}}_{T,i}^{(n)}(\ep)
\Big\vert_{n_F\gg 1}
\right]
=
\frac{\as}{\pi}\, C_i \, c(\ep)
\left(
\frac{1}{1+\frac{\as}{\pi} T_R \frac{5}{9} n_F \, k(\ep)} 
\right)
\,,
\end{equation}
where the function $k(\ep)$ takes the form
\begin{equation}
k(\ep) = 
-\frac{3}{5 \ep }+\frac{9 e^{\ep \gamma_E} (1-\ep ) \Gamma (1+\ep) \Gamma^2 (1-\ep )}{5 \ep  (3-2 \ep ) \Gamma (2-2 \ep )}
= 1 + \left(
\frac{28}{15}-\frac{\pi^2}{20}
\right) \ep + {\cal O}(\ep^2)\,,
\end{equation}
and the coefficients in the expansion are given by
\begin{equation}
{\widetilde\mathscr{A}}_{T,i}^{(n)}(\ep)
\Big\vert_{n_F\gg 1}
= C_i \, c(\ep)
\left(
- \frac{5}{9}\,T_R n_F
\right)^{n-1}
\left[ k(\ep) \right]^{n-1}
 ,
\hspace*{1cm} n \geq 1 \,.
\end{equation}
We note that starting from Eq.~(\ref{eq:ATallorders}) it is possible to take the conformal limit by setting $\epsilon=- \as \bnot^{(n_F)} $ and we obtain the following
all-order result\begin{equation}
{\widetilde\mathscr{A}}_{T, i}(\as; \ep=-\as \bnot )
\Big\vert_{n_F\gg 1}= {\frac{C_i \, \as}{\pi}}  \frac{\Gamma (4+2 \as \bnot^{(n_F)})}{6\, \Gamma (1-\as
    \bnot^{(n_F)}) \Gamma (1+ \as \bnot^{(n_F)}) \Gamma^2 (2+\as
    \bnot^{(n_F)})}\,,
\end{equation}
which is exactly the all-order expression for the cusp anomalous dimension $A_i(\as)$ in the large $n_F$ limit as obtained in Ref.~\cite{Beneke:1995pq}. 

In the case with $p=1$, 
we use 
Eq.~(\ref{eq:large_nf_tp}) and we find 
the following results for the perturbative coefficients of the soft coupling:
\begin{equation}\label{eq:A0coeff}
{\widetilde\mathscr{A}}_{0,i}^{(n)}(\ep)
\Big\vert_{n_F\gg 1}
\!\!\!\!\!= C_i \, c(\ep) \!
\left(T_R \frac{n_F}{3\ep}\right)^{n-1}
\sum_{k=0}^{n-1} \binom{n-1}{k} \!\!
\left[
-6 e^{\ep \gamma_E}
\frac{\Gamma^2(2-\ep)\Gamma(1+\ep)}{\Gamma(4-2\ep)}
\right]^k \!\!\!
\frac{\Gamma(1-\ep)}{\Gamma(1+k \ep)\Gamma(1-(k+1) \ep)}\,.
\end{equation}
The equivalence between Eq.~(\ref{eq:A0coeff})
and Eq.~(\ref{eq:large_nf_tp}) (for $p=1$)
can be proven
by performing the sum over $n$ in 
Eq.~(\ref{eq:A0coeff}) and exchanging the order of the $k$ and $n$ sums.
Considering the coupling 
${\widetilde\mathscr{A}}_{0,i}$ to all
orders in $\as$ and for generic values of $\ep$,
we are not able to obtain expressions that are more compact than Eqs.~(\ref{eq:large_nf_tp})
and (\ref{eq:A0coeff}).

For $\ep=0$ we can write very compact results for both $p=0$ and $p=1$.
In the case of ${\mathscr{A}}_{T,i}$ this is trivially obtained by setting $\ep=0$ in Eq.~(\ref{eq:ATallorders}):
\begin{equation}
\label{eq:ATallordersfourd}
\sum_{n \geq 1}
\left[
\left(
\frac{\as}{\pi}
\right)^n
{\mathscr{A}}_{T,i}^{(n)}
\Big\vert_{n_F\gg 1}
\right]
=
\frac{\as}{\pi}\, C_i
\left(
\frac{1}{1+\frac{\as}{\pi} \frac{5}{9} T_R n_F}
\right)\,.
\end{equation}
In the case of ${\mathscr{A}}_{0,i}$, we find the following result:
\begin{equation}\label{eq:A0allorders}
\sum_{n \geq 1}
\left[
\left(
\frac{\as}{\pi}
\right)^n
{\mathscr{A}}_{0,i}^{(n)}
\Big\vert_{n_F\gg 1}
\right]
=
\frac{\as}{\pi}\, C_i
\left[
\frac{3}{\as\,T_R n_F}
\arctan\left(
\frac{\as\,T_R n_F}{3}
\frac{1}{1+\frac{\as}{\pi} \frac{5}{9} T_R n_F}
\right) 
\right]\,.
\end{equation}
The derivation of Eq.~(\ref{eq:A0allorders})
is rather cumbersome and not particularly enlightening and therefore, for the sake of brevity, it is not reported here. However, starting  from Eq.~(\ref{eq:large_nf_tp}) it is rather straightforward, after expanding in $\as$ and $\ep$, to check the validity of Eq.~(\ref{eq:A0allorders}) to an arbitrary large power of the strong coupling.

The coefficients of the perturbative expansion 
of ${\mathscr{A}}_{T,i}$ and ${\mathscr{A}}_{0,i}$
can be obtained by computing the derivatives of 
Eqs.~(\ref{eq:ATallordersfourd}) and (\ref{eq:A0allorders})
with respect to $\as$. After some manipulation, they can be written as follows
\begin{equation}
{\mathscr{A}}_{T,i}^{(n)}
\Big\vert_{n_F\gg 1}
= C_i
\left(
- \frac{5}{9}\,T_R n_F
\right)^{n-1} ,
\hspace*{1cm} n \geq 1 \,.
\end{equation}
\begin{align}\label{eq:A0_all_orders}
{\mathscr{A}}_{0,i}^{(n)}
\Big\vert_{n_F\gg 1}
&= C_i
\left(
- \frac{5}{9}\,T_R n_F
\right)^{n-1}
\left(
1+\frac{9 \pi^2}{25}
\right)^{n/2}
\frac{5}{3 \pi} \frac{1}{n}
\sin \left[
n \arctan\left(
\frac{3\pi}{5}
\right)
\right] = \\[1ex]
&= C_i
\left(
- \frac{5}{9}\,T_R n_F
\right)^{n-1}
\frac{5i}{6\pi n}
\left[
\left(
1-\frac{3\pi i}{5}
\right)^n
-
\left(
1+\frac{3\pi i}{5}
\right)^n
\right]
, \nn
\hspace*{1cm} n \geq 1 \,.
\end{align}
The first few coefficients of Eq.~(\ref{eq:A0_all_orders}) read
\begin{align}
{\mathscr{A}}_{0,i}^{(1)}\Big\vert_{n_F\gg 1} &= C_i \,, \\
{\mathscr{A}}_{0,i}^{(2)}\Big\vert_{n_F\gg 1} &= C_i \left(
- \frac{5}{9}\,T_R n_F
\right)\,, \\
{\mathscr{A}}_{0,i}^{(3)}\Big\vert_{n_F\gg 1} &= C_i \left(
- \frac{5}{9}\,T_R n_F
\right)^{2}\left(1 - \frac{3\pi^2}{25}\right)\,,\\
{\mathscr{A}}_{0,i}^{(4)}\Big\vert_{n_F\gg 1} &= C_i \left(
- \frac{5}{9}\,T_R n_F
\right)^{3}\left(1 - \frac{9\pi^2}{25}\right)\,,\\
{\mathscr{A}}_{0,i}^{(5)}\Big\vert_{n_F\gg 1} &= C_i \left(
- \frac{5}{9}\,T_R n_F
\right)^{4}\left(1 - \frac{18\pi^2}{25} + \frac{81\pi^4}{3125} \right)\,,
\end{align}
and the results for $n\le 4$ agree with the large-$n_F$ limit of the corresponding exact result~\cite{Catani:2019rvy}.

\section{Summary}
\label{sec:summa}

In this paper we have studied extensions of the soft-gluon effective coupling in the context of soft-gluon resummation
beyond NLL accuracy.
Up to NLL accuracy the intensity of soft radiation is controlled through the CMW coupling \cite{Catani:1990rr} and is universal, i.e., the soft coupling takes the same form in the resummation program for different hard-scattering observables. Beyond NLL order there is no unique extension of the soft-gluon coupling. 
Starting from the probability density $w_i(k;\epsilon)$ of correlated soft emission,
we have introduced an entire class of soft-gluon effective couplings
that are specified by a scale that depends on the transverse momentum and the transverse mass
of the inclusive soft radiation.
These couplings are relevant for resummed QCD calculations of different hard-scattering observables.

We have shown that all these couplings are equal at the conformal point where the $d$-dimensional QCD $\beta$ function vanishes,
thereby extending the result of Ref.~\cite{Catani:2019rvy}.
We have presented explicit results for the probability density $w_i(k;\epsilon)$
and the soft couplings at the second order in the QCD coupling $\as$ in $d$ space-time dimensions.
We have also shown that at this perturbative order all the soft couplings are equal to the CMW coupling in the physical four-dimensional space-time.
In $d=4$ dimensions, we have derived an explicit
relation between the soft couplings at 
${\cal O}(\as^3)$. This relation and the 
${\cal O}(\as^3)$ results for $\AT$ 
\cite{Banfi:2018mcq, Catani:2019rvy} and $\Ao$ 
\cite{Catani:2019rvy}
directly give the 
${\cal O}(\as^3)$ result for any soft coupling
of the class that we considered in this paper.

Finally, we have computed the all-order structure of the probability density $w_i(k,\epsilon)$ in the large-$n_F$ limit and we have presented
explicit results for the soft couplings
$\AT(\as)$ and $\Ao(\as)$ at large $n_F$ to all perturbative orders.
We have also shown that, as expected, our large-$n_F$ results at the conformal point are consistent with
the known structure of the cusp anomalous dimension \cite{Beneke:1995pq}.

\noindent {\bf Acknowledgements}. This work is supported in part by the Swiss National Science Foundation (SNSF) under contract 200020$\_$188464. One of us (MG) would like to thank the CERN Theoretical Physics Department for the kind hospitality and financial support during the completion of this work.

\bibliography{biblio}

\end{document}